\documentclass[multphys,vecphys]{svmult}

\usepackage{makeidx}         
\usepackage{graphicx}        
\usepackage{multicol}        
\usepackage[bottom]{footmisc}
\usepackage{cite}
\usepackage{amsmath}
\usepackage{amssymb}

\makeindex             

\begin{document}

\title*{Mixing, Ergodicity and the Fluctuation-Dissipation Theorem in complex systems}
\titlerunning{Mixing, Ergodicity...} 
\author{Mendeli H. Vainstein \and Ismael V. L. Costa  \and F. A. Oliveira}
\institute{Institute of Physics and International Center of Condensed
Matter Physics, University of Bras\'{\i}lia, CP 04513, 70919-970,
Bras\'{\i}lia-DF, Brazil \texttt{fao@fis.unb.br}}
%
%
\maketitle

ABSTRACT:
Complex systems such as glasses, gels, granular materials, and systems far from equilibrium exhibit violation of the ergodic hypothesis (EH) and of the fluctuation-dissipation theorem (FDT). 
 Recent investigations  in systems with memory~\cite{Costa03} have established a hierarchical connection between mixing, the EH and the FDT. They have shown that a failure of the mixing condition (MC) will lead to the subsequent failures of  the EH and of the FDT. Another important point is that such violations are not limited to complex systems:  simple systems may also display this feature. Results from such systems are analytical and obviously easier to understand than those obtained in complex structures, where a large number of competing phenomena are present.
 In this work, we review some important requirements for the validity of the FDT and its connection with mixing, the EH and anomalous diffusion in one-dimensional systems. 
We show that when  the FDT fails, an out-of-equilibrium  system relaxes to an effective temperature different from that of the heat reservoir. This effective temperature is a signature of metastability found in many complex systems such as spin-glasses and granular materials.

\section{Introduction}
\label{sec.1}

Since its formulation by Boltzmann~\cite{Boltzmann74}, the EH has called the attention of mathematical physicists and chemists. In the last century, a branch of the mathematics dedicated to its study has been developed. However, most of its results are accessible only to the specialist.
%
 On the other hand, the FDT has played a central role~\cite{Kubo66,Kubo91} in nonequilibrium statistical mechanics in the linear response regime (NESML). It gained such importance that Kubo proposed a complete formulation of the NESML based on it~\cite{Kubo91}. Since the FDT is directly related to relaxation processes, its more empirical character has caught the attention of experimentalists and most of the discussion about its validity has remained in the hands of theoretical physicists and chemists, instead of mathematicians.   
        
 A necessary requirement for the validity of the FDT is that the time-dependent dynamical variables be well defined at equilibrium. The presence of nonlinear effects or far from equilibrium dynamics may lead to situations where the FDT does not hold~\cite{Costa03,Bellon05}, the aging process in spin-glass systems being a good example~\cite{Parisi97,Kauzmann48,Santamaria-Holek01,Ricci-Tersenghi00,Exartier00,Grigera99}.

Most of experimental situations in which the EH and the FDT are violated happen in complex structures. Nevertheless, we show here simple situations where those violations appear. Their explicit condition and simplicity allows the judgment of  non-specialists in the subject. This is important because it opens the possibility that more complex structures can be investigated on a solid basis.

This work is organized as follows: In this section, we shall define  the EH, the MC, and other main concepts and ideas to be discussed throughout this work.
 In Section~\ref{sec.diffusion}, we outline some historical achievements in the study of diffusion and introduce the FDT. We discuss reaction rates in Sec.~\ref{sec.rate}, before introducing, in Sec.~\ref{sec.memory}, the concept of memory, i.e., we discuss  a system governed by a Generalized Langevin's Equation (GLE) and we show how to obtain anomalous diffusion. Next, we discuss random walks,  fractional derivatives and their connection to the GLE in Sec.~\ref{sec.random}. We then continue by defining in a clear manner the noise in Sec.~\ref{sec.noise}, and its connection with memory and correlation functions. After that, we discuss reversibility in Sec.~\ref{sec.reversibility}. 
 In Section~\ref{sec.violation}, we discuss the main issue of this review, which is the interconnection between the MC, the EH, and the FDT, and we show under what conditions they fail. Examples of such violations are given in Sec.~\ref{sec.ballistic}, where we study ballistic motion.
In Section~\ref{sec.shape}, we introduce some speculative topics on the forefront of physical research; the ``skeptic reader'' can skip it. Finally, we introduce a conjecture in Sec.~\ref{sec.conjecture}  and conclude the paper in Sec.~\ref{sec.conclusion}.

\begin{figure}
\centering
\includegraphics[height=7cm,width=5cm,angle=270]{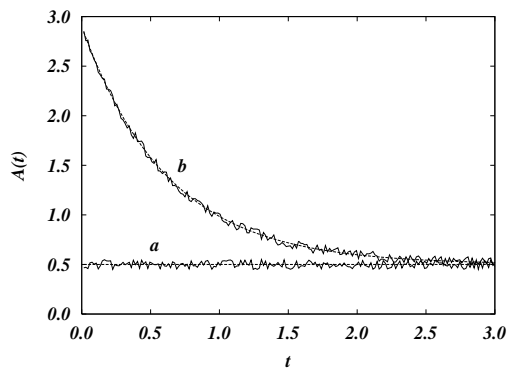}
\caption{Evolution of a dynamical stochastic variable.}
\label{fig.1} 
\end{figure}

Let us start by  considering the evolution of a dynamical stochastic variable $A(t)$ see Fig.~\ref{fig.1}. The variable could be either at equilibrium, Fig.~\ref{fig.1}$(a)$, or approaching it, Fig.~\ref{fig.1}$(b)$. The ensemble average $\langle G(A)\rangle $ of any function $G(A)$ is defined as
\begin{equation}
\label{ensemble}
\langle G(A)\rangle =\int_\Omega \exp(-\beta E(A))G(A) d\Omega,
\end{equation}  
where $\beta^{-1}= k_BT$ is the inverse temperature, $E(A)$ are the energies, and the integration is performed  over all the accessible states of the phase space $\Omega$. 
From that, it is possible to define a correlation function as
\begin{equation}
\label{Cv}
C_A(t)=\langle A(t)A(0)\rangle .
\end{equation}
 For an exponential decay of the correlation function, it is possible to associate a relaxation time $\tau$, 
which is larger than the typical time for a fluctuation $\Delta t$. One can also define a time average as
                                                                               
\begin{equation}
\overline{A(t)}=\frac{1}{T}\int_{-T/2}^{T/2}A(t+t')dt'
\label{time}
\end{equation}
For  $\Delta t\ll T\ll \tau$, the average will produce the continuous line in Fig.~\ref{fig.1}, i.e., it will wash out the fluctuations which we measure with a sensitive probe. For times $T \gg \tau$, the EH reads
                                                                               
\begin{equation}
\overline{A(t)}= \langle A(t)\rangle .
\label{EH}
\end{equation}
In simple words: \emph{given enough time, the system will reach every accessible state}, and a time average will be equal to an ensemble average. 
The proof of the EH has been the Holy Grail of  statistical mechanics. One expects it will hold for macroscopic systems at equilibrium, 
Fig.~\ref{fig.1}$(a)$, and for small deviations from equilibrium; on the other hand, it will not hold for curve $(b)$, although it is expected that the system in situation $(b)$ will be driven to equilibrium for longer times. However, a general proof of the EH  is still missing.
 The concept of ``far from equilibrium'' is itself sometimes misleading, since it depends not only on the initial conditions, but also on the possible trajectories the system may follow~\cite{Costa03,Costa04}.
The way a system approaches an equilibrium is crucial for these definitions.

 The ``mixing property of a physical system'' or mixing condition (MC) can be stated as
\begin{equation}
\lim_{t \rightarrow \infty }R(t)=0,
\label{MC}
\end{equation}
where we have defined the normalized quantity $R(t)=C_A(t)/C_A(0)$. 
The MC tells us that after a long time $t\gg \tau$, we do not expect that $A(t)$ will remember its initial value $A(0)$. 

\section{Diffusion}
\label{sec.diffusion}

 Diffusion is one of the most fundamental mechanisms for transport of energy, mass and information; it is a main process for a system  to reach uniformity and equilibrium, and has therefore been the focus of extensive research in many different disciplines of natural science. For almost two hundred years, it has caught the attention of the scientific community. 
The  famous observations by Robert Brown~\cite{Brown28,Brown28a} of the erratic trajectories of pollen opened a new world for experimental and theoretical studies in what was named Brownian motion.
 As a biologist, Brown first assumed he had discovered the basic essence of life, an idea to be expected from a  man dedicated to biology. Nevertheless, he reproduced the experiments with non-organic material and observed the same erratic motion. 
Brown then concluded that this was due to the motion of matter. Considering his aim, the second conclusion is not only more difficult; it is a highly advanced and honest conclusion. 
Unfortunately, most books are unfair with Brown, in not mentioning his subsequent experiments. 
 
At the centennial celebration of the Einstein miraculous year, one could be easily driven to the conclusion that most of the diffusive phenomena are well understood today. 
However, if we ask simple questions such as ``How do spin waves diffuse in a Heisenberg system with correlated disorder?'', ``How do electrons behave in an irregular lattice?'', or   ``How does a ratchet device work?'', it takes a short time to realize that these unanswered problems are related to diffusion.

For instance, when we flip a spin in the ground state of a ferromagnetic chain, the principle of equal \emph{a priori} probability for the accessible states tells us that somehow the energy due to this disturbance will not remain localized in a single state.
 However, it does not say whether or not the system will support a spin wave,  whether the wave propagates, and how it propagates in the affirmative case. To answer this kind of question, we  usually need to go on into specific calculations. Our aim is to try to understand the general character of diffusion and, hopefully, to  classify it prior to extensive calculations.

At the beginning of the twentieth century, much important research was dedicated to the irregular motion of microscopic particles dispersed in a fluid, namely Brownian motion. Despite being irregular, the motion reveals some regularity when analyzed statistically. 
The main observed quantity was the mean square displacement  $\langle x^2(t)\rangle $ of the particles, which evolves linearly with time.

 Einstein's basic idea was to explain the Brownian motion by going beyond  thermodynamics  and into kinetic theory. He considered single spherical particles of mass $m$ and radius $a$ suspended in a liquid of viscosity $\eta$, and obtained~\cite{Einstein56}
                                                                                
\begin{equation}
\lim _{t\rightarrow \infty }\langle  x^{2}(t)\rangle =2Dt.
\label{X2}
\end{equation}
As usual, by infinite time, we mean a time larger than the maximum relaxation time of the process. The diffusion constant $D$ was given by
\begin{equation}
D=\frac{RT}{6\pi N_a a \eta}=\frac{RT}{mN_a\gamma}=\frac{RT\mu}{N_a},
\label{D}
\end{equation}
where $R$ is the gas constant, $N_a$ the Avogadro number, $\mu$ the mobility, and $\gamma$ the friction the particle feels in the fluid.

 Equation~(\ref{X2}) was a major achievement; the linear relation with time was confirmed and an expression for the diffusion constant was obtained.
 The last form for $D$ in Eq.~(\ref{D}) establishes a connection to the mobility, which is fundamental for the study of conductivity and transport.
 It was then possible to check the theory with the data available at the time.
 If one knows $D$, it is possible to make an estimation of the Avogadro number $N_a$. Future works helped establish the Boltzmann constant, $k_B=R/N_a$, as a new fundamental constant. Moreover, estimation of the size of sugar molecules dissolved in water  became possible. Besides that, the frequency-dependent diffusion constant $\widetilde{D}(\omega)$ can be directly associated with the conductivity by the relation~\cite{Dyre00}
\begin{equation}
 \widetilde{\sigma}(\omega)=\frac{ne^2}{k_BT}\widetilde{D}(\omega),
\label{Domega}
\end{equation}
where $e$ is the carrier charge and $n$ the carrier density.

 The series of Einstein articles about diffusion~\cite{Einstein56} together with the work of Smoluchowski~\cite{Smoluchowski06} paved the way for  the modern theory of Brownian motion.

The next step forward was taken by Langevin.
 In order to  describe the motion of a particle immersed in a fluid, in 1908, Langevin~\cite{Langevin08} proposed the equation
\begin{equation}
m\frac{dv(t)}{dt}=-m\gamma v(t)+f(t).
\label{L}
\end{equation}
 where $f(t)$ is a stochastic force subject to the conditions  $\langle  f(t)v(0)\rangle =0$, $\langle  f(t)\rangle =0$ and $\langle  f(t)f(t')\rangle= \Lambda \delta(t-t')$. 
 If we solve Eq.~(\ref{L}) and by using the equipartition theorem we impose $\langle  v^2(t \rightarrow \infty) \rangle=k_BT/m$, we obtain the proportionality constant $\Lambda =2mk_BT \gamma$ and write

\begin{equation}
\langle f(t)f(t')\rangle=2mk_BT\gamma\delta(t-t').
\label{FDT0}
\end{equation}
This last relation is known as the fluctuation-dissipation theorem (FDT). 
 Although the Einstein diffusion constant contains implicitly the relation between fluctuation and dissipation, 
in Langevin's formulation, it acquires the importance of a basic theorem.

 From Eq.~(\ref{L}) and the above conditions, one obtains the velocity-velocity correlation function
\begin{equation}
\label{Cv1}
   C_v(t)=  \langle v(t+t')v(t')\rangle=(k_BT/m)  \exp(-\gamma t).
\end{equation}
 The correlation function $C_v(t)$, or $R(t)=\exp(-\gamma t)$, will satisfy Eq.~(\ref{MC}), the MC, with a relaxation time $\tau=\gamma^{-1}$.
 This exponential decay from the initial conditions is the expected form for the MC. Now, it is possible to obtain the mean square displacement as
\begin{equation}
\label{X22}
\langle x^2(t\gg \tau) \rangle= \int_0^tdt'\int_0^t dt''\langle v(t')v(t'')\rangle =2Dt,
\end{equation}
where, in the last step, we used the Kubo formula
\begin{equation}
\label{Kubo}
 D=\int_0^{\infty}C_v(t)dt.
\end{equation}
The Kubo formula, together  with Eq.~(\ref{Cv1}), reproduces Einstein's results.

There are many reasons why one should always look back to Langevin's work, the first one being that it focuses attention on the motion of a particle, which is very intuitive for any physicist.
 Second, it combines the old Newtonian deterministic approach with the new ``uncertainty'' of the stochastic force $f(t)$. 
The breaking of the atomic forces in two parts: a fast changing force $f(t)$ with time scale $\Delta t$, and the slow friction force with time scale $\tau$, introduces a large simplification which facilitates understanding and computer simulation.
 Consequently, the use of the Langevin equation, and of its generalization (Sec.~\ref{sec.memory}), is still very active, having been applied successfully to the study of many different systems such as the dynamics of dipolar~\cite{Toussaint04} and polymeric chains~\cite{Oliveira94,Oliveira96,Oliveira98a,Maroja01}, metallic liquids~\cite{Rahman62},  Lennard Jones liquids~\cite{Yulmetyev03}, 
diffusion in periodic potentials~\cite{Sancho04}, ratchet devices~\cite{Bao03, Bao03a}, and synchronization~\cite{Ciesla01}, only to name a few.  
Finally, it established explicitly for the first time the connection between fluctuation and dissipation, the FDT, which remains a major theorem of statistical mechanics.

The Langevin equation, however, presents some limitations: (a) It is a classical formalism; (b) It has uncorrelated noise with only two time scales $\Delta t$ and $\tau$, whereas  a complex system has in general many time scales;
(c) We cannot make any predictions for times shorter than $\Delta t$; 
(d) It predicts only normal diffusion.

  A  quantum formulation of the FDT has been put forward by Callen and Welton~\cite{Callen51}. Following their work, much research has been done in the field, with many attempts at generalization~\cite{Parisi97,Belyi04}. We shall focus our attention on the Kubo FDT~\cite{Kubo57,Kubo57a,Kubo91}, or the so-called second FDT, since it is more useful to the study of diffusion; see Sec.~\ref{sec.memory}. 

\section{Reaction Rates}
\label{sec.rate}

\begin{figure}
\centering
\includegraphics[height=7cm,width=5cm,angle=270]{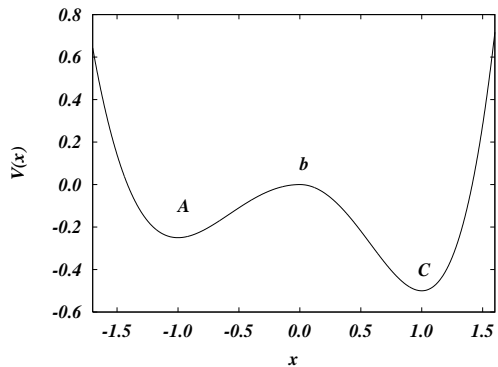}
\caption{Potential with two minima.}
\label{fig.2} 
\end{figure}

Diffusion may be considered the simplest problem of nonequilibrium statistical mechanics; however, if one considers a particle moving in an irregular media, it is quite probable that the particle will  be affected by some potential and, in moving from point $A$ to point $C$, will have to cross a potential barrier at point $b$, see Fig.~\ref{fig.2}. 
This is one of the oldest problem in statistical mechanics and was first mentioned by Arrhenius~\cite{Arrhenius89}.  Arrhenius proposed that the rate of particles crossing the potential barrier will be given by
\begin{equation}
k=\nu_0 \exp(-\beta E_b),
\label{Ahr}
\end{equation}                                                                                
where $\nu_0$ is the attempt frequency and $E_b$ is the barrier height. Arrhenius associated the attempt frequency to the vibration frequency at the well bottom.

The issue stood still until Kramers~\cite{Kramers40} re-addressed it with a more accurate analysis,
 in such way that the problem is known today as the Kramers problem. He gave an expression for the attempt frequency, which depends                                                        on the curvature at the top of the potential barrier and on the friction. The Kramers theory of reaction rates has many applications in biology, chemistry and physics. For a review, see~\cite{Hanggi90}.

In a recent work, Sancho \emph{et al.}~\cite{Sancho04} studied the diffusion of particles in a periodic potential. They simulated the diffusion using Langevin's equation and  used Kramers reaction rate to analyze the jumping between successive wells. A good agreement between theory and the simulation was obtained. This is a nice exhibition of the relation between transport processes and reaction rates.

For systems with memory~\cite{Mori65,Grote80,Pollak90}, the Kramers problem suffers  the same restrictions as those of diffusion; worse,
 some of the restrictions are not even as explicit as in diffusion~\cite{Oliveira98}. In short, the reaction rate theory will be enriched by a better understanding of diffusion.

\section{Complex Systems have Memory}
\label{sec.memory}

If one asks ``What makes a system complex?'', a few concepts will come to mind:  a large number of degrees of freedom, nonlinearity and memory. However, if we ask for a definition of  memory, only a few answers will be precise.
We shall use here the concept of memory introduced by Mori in his seminal paper~\cite{Mori65}, where he used a method of projection operators which has many advantages: it allows the treatment of  quantum systems, it is not empirical, it has time correlation, and it is a non-Markovian formulation with an explicit definition of memory.

Forty years after Mori's work, many fundamental concepts and methods
have been developed~\cite{Zwanzig01,Kubo66,Kubo91,Mori65,Evans90,Lee83a,Lee00,Lee01,Marcus60,Nakajima58}, 
which allowed a generalization of the Langevin formalism and the elimination of most of its limitations (see items  (a) to (d) at the end of Sec.~\ref{sec.diffusion}).  The new  formalism  gives origin to a Generalized Langevin Equation (GLE) of the form
                                                                 \begin{equation}
\frac{dA(t)}{dt}=-\int _{0}^{t}\Pi (t-t')A(t')dt'+F(t),
\label{GLE}
\end{equation}
 where $F(t)$ is a stochastic noise subject to the conditions $\langle F(t)\rangle =0$,
$\langle F(t)A(0)\rangle =0$ and

\begin{equation}
C_{F}(t)=\langle F(t)F(0) \rangle= \langle A^{2}\rangle_{eq}\Pi(t).
\label{FDT}
\end{equation}
  Equation~(\ref{FDT}) is the 
Kubo FDT~\cite{Kubo57,Kubo57a,Kubo91}, being a generalization of Eq.~(\ref{FDT0}).
The memory, $\Pi (t)$, arises here explicitly and, in principle, it allows us to study a large number of correlated processes.
  Notice that now many time scales are possible within $\Pi(t)$; this is a natural condition for complexity.
 An equation for $C_A(t)$, or for the renormalized correlation function $R(t)$ is given by 
\begin{equation}
\frac{dR(t)}{dt}=-\int _{0}^{t}\Pi (t-t')R(t')dt',
\label{GR}
\end{equation}
where we have used the conditions $\langle A(0) F(t) \rangle=0$. The Laplace transform of this equation yields
\begin{equation}
\widetilde{R}(z)=\frac{1}{z+\widetilde{\Pi}(z)}.
\label{Rz}
\end{equation}
From here on, we shall use the tilde to indicate Laplace transforms.

Let us now define the variable $y(t)$ as
\begin{equation}
y(t)=\int _{0}^{t}A(t')dt',
\label{3}
\end{equation}
 with asymptotic behavior
\begin{equation}
\lim _{t\rightarrow \infty }\langle y^{2}(t)\rangle \sim t^{\alpha }.
\label{y2}
\end{equation}
 For normal diffusion, $\alpha =1$; we have subdiffusion for $\alpha <1$, and superdiffusion for $\alpha >1$. Notice that if $A(t)$ is the momentum of a particle, then $y(t)/m$ is its position.
                                                                                
It is very simple to show that this new formalism allows both normal and anomalous diffusion. Consider two examples: 
first, take 
\begin{equation}
\Pi(t)= 2\gamma \delta(t).
\end{equation}
 With this short range memory, we return to the normal Langevin's equation, Eq.~(\ref{L}), and obtain $\alpha=1$, i.e., normal diffusion.  Second, consider an  extremely long memory
\begin{equation}
\Pi(t)= \omega_0^2,
\label{HO}
\end{equation}
which gives a force of the form $-m\omega_0^2 y$, i.e., an harmonic oscillator which does  not exhibit diffusion at all, $\alpha=0$. Those are artificial but simple examples of how the memory determines diffusion.

Recently, Morgado \emph{et al.}~\cite{Morgado02}
obtained a general classification for anomalous diffusion. They used the time-dependent diffusion function
\begin{equation}
D(t) = \int_0^t C_v(t')dt',
\end{equation}
\begin{equation}
\lim _{t\rightarrow \infty } D(t)=\lim_{z \rightarrow 0}z\widetilde {D}(z)=\lim_{z \rightarrow 0} \widetilde{\Pi}^{-1}(z)=\lim_{t \rightarrow \infty} \frac{1}{\widetilde{\Pi}(1/t)}.
\label{D(t)}
\end{equation}
Here, we have used the final value theorem~\cite{Spiegel65}, and Eq.~(\ref{Rz}). 
Note that the long range  time behavior, i.e. the dynamics, is dominated by the  small frequencies. 
We expect that  $z \rightarrow 0$ faster than $\widetilde{\Gamma }(z) $.  
Consider now that 
\begin{equation}
  \widetilde{\Pi }(z\rightarrow 0) \sim z^{\nu}.
\label{Gaz}
\end{equation}
 With $t \sim z^{-1}$, we get
\begin{equation}
\alpha= \nu+1.
\label{alpha}
\end{equation}

This result is fundamental for the classification of diffusion. Let us take the previous examples: first, $\Pi(t)=2\gamma \delta(t)$ with Laplace transform $\widetilde{\Pi}(z)=\gamma$; here $\nu=0$, $\alpha=1$, 
and the diffusion is normal as expected. Second, from Eq.~(\ref{HO}), we get
$\widetilde{\Pi}(z)=\omega_0^2/z$, $\nu=-1$, and $\alpha=0$.

We are now in condition to discuss mixing.
Consider a system governed by a GLE. The asymptotic behavior
\begin{equation}
\lim _{t\rightarrow \infty } \Pi(t)=\lim_{z \rightarrow 0}z\widetilde {\Pi}(z),
\label{Galim}
\end{equation}
\begin{equation}
\lim _{t\rightarrow \infty }R(t)=\lim_{z \rightarrow 0}z\widetilde {R}(z)=\lim_{z \rightarrow 0}\frac{z}{z+\widetilde{\Pi}(z)},
\label{Rlim}
\end{equation}
and condition (\ref{Gaz}) show that mixing exists only for $-1<\nu<1$. In other words, the mixing condition is fulfilled between the limits of ballistic motion and harmonic oscillator motion. Close to the limits, one can expect problems.

Notice that for the memory given in Eq.~(\ref{HO}), we have the exact solution
\begin{equation}
\label{RHO}
R(t)=\cos(\omega_0t),
\end{equation} 
which obviously does not fulfil the MC, Eq.~(\ref{MC}).
Notice also that if Eq.~(\ref{Kubo}) converges, then  it is always possible to have an associated ``friction constant'' $\gamma$, even for correlated systems
\begin{equation}
\label{gamma}
\gamma^{-1} = \int_0^{\infty}R(t)dt.
\end{equation}
This definition can be plugged  into the diffusion formula (\ref{D}). This
explains why normal diffusion is so widely found in nature, even for processes we clearly know are strongly correlated~\cite{Morgado02}. For a long time, the only evidence of anomalous diffusion had been for subdiffusion processes due to trapping mechanisms~\cite{Dyre00} and hierarchical lattices~\cite{Alexander82}. However, superdiffusive motion has recently been studied both theoretically~\cite{Baskin04,Budini04} and experimentally~\cite{Ferreira91,Frank98,Bellani99,Bellani00,Poncharal02,Monte03} (see Sec.~\ref{sec.ballistic}). We also expect that research in the new nanoclay technology will produce anomalous diffusion~\cite{Bakk02}. Recent works on chain dynamics~\cite{Oliveira95,Oliveira98a,Toussaint04} show that the system dynamics may build up a memory. This ``casual'' result is explored in the conjecture described in Sec.~\ref{sec.conjecture}.

\section{Random Walk}
\label{sec.random}

   The study of random walks is somehow older than statistical mechanics and it has  produced many alternative ways to describe diffusive processes. Even before Einstein, the mathematical works of Bernoulli opened up the possibility of understanding fluctuations; for example one can arrive at Eq.~(\ref{X2}) by considering the famous drunk man problem, a starting point in many undergraduate texts~\cite{Salinas01}.
Besides that, the famous law of  large numbers, or $N^{-1/2}$, for the relative standard deviation of a variable suggested that if atoms exist they must be very small, with $N$ being very big, otherwise thermodynamics would not make sense. 
However, the famous expression for diffusion Eq.~(\ref{D}) was not obtained before Einstein; it has the basic information one needs to know about the nature of the process. That is a main difference between mathematics and physics. In this sense, the  Einstein diffusion constant was a stunning achievement. We shall call the attention to the famous Chandrasekhar review on stochastic  process~\cite{Chandrasekhar43}, 
which was an up-to-date article until the origin of the Mori formalism. It still remains  as a clear and concise review.

  By the end of the nineteen century, the works of Lord Rayleigh~\cite{Rayleigh64} on random flights allowed to understand scattering in random directions, one of the most famous application of  which  was to explain the blue color of the sky, due to light scattering by impurities in the atmosphere.
 A modern formulation shows the latter approach is incorrect and that light scattering is associated with fluctuations in the dielectric constant~\cite{Loudon00}.  After the creation of the laser, light scattering became itself a large field of research.~\cite{Scalabrin77,Rousseau81}

 Let us consider here a simple random walk analogy to Langevin's work. Consider a set of $N_0$ particles with initial velocities equal to zero.The particles are subject to a random force $\pm f_0$ at each time interval $\Delta t$ with equal probability. After $N$ time steps, the average velocity will be $\langle v(t)\rangle=0$ and the average squared velocity
\begin{equation}
\langle v^2(t)\rangle= N f_0^2 \Delta t^2/m^2.
\end{equation}
We can see from the previous expression that the kinetic energy grows linearly with time, $t=N \Delta t$, i.e., the random force acts as a pump of energy.      Such a simple model does not represent a physical process, since it contradicts the kinetic theory, $\langle v^2(t\rightarrow \infty) \rangle =k_BT/m$. To make it more realistic,  we add a dissipative force $-m\gamma v(t)$, and impose the balance of energy to obtain $f_0^2= 2mk_BT\gamma/ \Delta t$.
 Notice that if we define the Dirac delta function as the limit $\Delta t \rightarrow 0$ of $\delta(t)=1/\Delta t$ for $- \Delta t/2 \le t \le \Delta t/2$, and $0$, otherwise, we recover the FDT, Eq.~(\ref{FDT0}). The FDT is nothing more than a detailed balance condition; 
it is a guarantee that the dispersed particles will reach thermal equilibrium after a reasonable time.

    In the same way, there are alternative ways to describe  anomalous diffusion besides the GLE.  One proposal is to use the continuous random walk, which can be mathematically described by fractional derivatives~\cite{Chaves98,Metzler00,Metzler04}. 
The fractional derivative of a function $f(x,t)$ can be defined as
\begin{equation}
_0D_t^{1-\alpha}f(x,t)=\frac{1}{\Gamma(\alpha)}\frac{\partial}{\partial t}\int_0^t dt'\frac{f(x,t')}{(t-t')^{1-\alpha}},
\label{fd}
\end{equation}
where $\Gamma(x)$ is the Gamma function. Equation~(\ref{fd}) is a natural generalization of the derivative of a complex variable using the residue theorem. 
The nonlocal character of the fractional derivative is the same as that of the memory. Therefore, it is quite natural that they yield the same results as the GLE. 
Indeed, it is possible to obtain a fractional Fokker-Planck equation (FFPE) of the form 
\begin{equation}
\label{FFPE}
\frac{\partial f}{\partial t}=\,  _0D_t^{1-\alpha} \left[ \frac{\partial}{\partial x}\frac{V'(x)}{m \eta_{\alpha}}+ K_{\alpha}\frac{\partial^2}{\partial x^2}\right]f(x,t) ,
\end{equation}
to address the problem of subdiffusion~\cite{Metzler00} and to obtain a relation similar to Eq.~(\ref{alpha}). 

 The generalization for superdiffusion was recently discussed in a few articles ( \cite{Metzler04}, and references therein).
Fractional derivatives are a very compact way to obtain results; however, there are some points one should bear in mind.

First, we assume \emph{a priori} a fractional geometry and as a result  we obtain fractional diffusion; that appears to be a circular argument. From the GLE, that comes naturally from the memory, or from the noise; see Eq.~(\ref{Gamma}). 

Second, if $x(t)$ gives the time evolution for a particle position, we know precisely what $ _0D^{\mu}_t x(t)$ means only for $\mu=1,2,\ldots$ However, no one has an idea of what it means for $\mu=0.51$, or for any non-integer value.

Finally, natural solutions for the fractional derivatives are the L\'evy functions. Unfortunately, these yield an infinite mean square displacement, which is not a good physical result. 
The L\'evy distributions, $\phi_{\mu}(x)$, are very popular because they fulfill the generalized  central limit theorem (GCLT)~\cite{Levy37}       

\begin{equation}
 \phi_{\mu}(x)=\int \phi_{\mu}(x-x')\phi_{\mu}(x')dx'.
\end{equation}                                              
Moreover, recently~\cite{Oliveira00,Oliveira01a}, it has been shown that the GCLT   represents the first uncorrelated term in a renormalization process.
 Correlations, such as those one expects to find in anomalous diffusion, will cause the deviation of the studied variables from the GCLT. Recently, Figueiredo \emph{et al.}~\cite{Figueiredo03,Figueiredo03a,Figueiredo04,Figueiredo04a}, have proposed some theorems on the limit sum of stochastic variables without making the classical assumptions of the GCLT. They developed a general formalism to explain the non-convergence (or the slow convergence) to the Gaussian distribution. 
With this, they have explained the origin of the self-similar property that appears in real economics time series data. 
They have also explained how autocorrelations (linear and nonlinear) can be considered as a source of truncated L\'evy flights.
 The asymmetry they have found in their distributions are similar to those
found in relaxation in supercooled liquids and in the height distributions
in the etching of a crystalline solid~\cite{Mello01,Reis03,Reis04}. The experiments of
Monte \emph{et al.}~\cite{Monte00,Monte02} show both asymmetric and superdiffusive behavior.

\section{Noise}
\label{sec.noise}

   A fundamental aspect of stochastic processes is the noise.  
A stochastic generalized force $F(t)$ can be decomposed into a set of harmonic oscillators of the form

\begin{equation}
\label{Noise}
 F(t)= \langle A^2\rangle_{eq}^{1/2} \int \rho(\omega)^{1/2}\cos(\omega t +\phi(\omega))d\omega
\end{equation}
Here,  $0<\phi(\omega)<2\pi$ is the random phase. From the FDT, it follows that
 \begin{equation}
\label{Gamma}
\Pi (t)=\int \rho (\omega )\cos (\omega t)d\omega,
\end{equation}
where $\rho(\omega)$ is the noise density of states (NDS).
     Now, we take the Laplace transform of Eq.~(\ref{Gamma}) to obtain
\begin{equation}
\label{gamma2}                          
\gamma =\lim_{z \rightarrow 0}\widetilde{\Pi}(z) = \frac{\pi}{2}\rho(0).
\end{equation}
This is another relevant result. The friction is equal  to  the noise density of states at the origin. This shows how the lower modes determine the type of diffusion. 
A system which has a finite friction  presents normal diffusion, since its NDS is finite at the origin.
 Subdiffusion will have an infinite  friction and an infinite  NDS. 
Finally, superdiffusion has a null friction. To obtain superdiffusion for a null friction is a very intuitive and appealing concept. A null NDS tells us that the lower modes do not relax and the process has ``less interference'', or, in Langevin's language, ``weak collisions''. 

Consider now the colored noise
\begin{equation}
\label{N1}
\rho(\omega) =
 \begin {cases}
 \frac{2 \gamma_0}{\pi}(\frac{\omega}{\omega_D})^{\beta} & \text{, if }\omega<\omega_{D}\\
 0 & \text{, otherwise}.
\end{cases}
\end{equation}
Here, $\omega_D$ is a Debye cutoff frequency. This kind of noise has been used by Caldeira and Leggett in quantum dissipative systems~\cite{Caldeira83}. If we plug this noise into Eq.~(\ref{Gamma}), take its  Laplace transform, and then  the limit of small $z$, we obtain the exponent~\cite{Vainstein05a} from Eq.~(\ref{Gaz})

\begin{equation}
\label{beta}
\nu  =
  \begin{cases}
 \beta &\text{, if } \beta < 1\\
 1 & \text{, otherwise}.
\end{cases}
\end{equation}

For most of the cases, the exponent of the NDS for low frequencies will be the same as that of the Laplace transform of the memory for small $z$. 
 Equation~(\ref{beta}) shows that  $\alpha \le 2$ and, consequently, the motion is limited by the ballistic motion. Ballistic motion appears to be a limit of this kind of GLE, see Sec.~\ref{sec.violation}. 
Notice as well that for $\nu=0$, we get $\gamma=\gamma_0$, from Eq.~(\ref{gamma2}).

We shall consider here another possibility. Let the noise be 

\begin{equation}
\rho(\omega) = 
  \begin{cases}
 \frac{2\gamma_0}{\pi} & \text{, if } \omega_{1}<\omega<\omega_{2}\\
 0 & \text{, otherwise.}
\end{cases}
\end{equation}
For $\omega _{1}=0$, we have the Debye density of states for a thermal
noise composed of acoustic phonons. Thus, for $\omega _{1}=0$ we
have normal diffusion and for any $\omega _{1}\neq 0$ we have superdiffusion.
This NDS is the difference between two Ornstein-Uhlenbeck processes and is a simple way  to produce  ballistic diffusion~\cite{Morgado02,Costa03}. 
Since there is a window, $0<\omega<\omega_1$, where there is no fluctuation of the modes, this introduces a  very practical mechanism to control simulations.
 This kind of noise seems more appropriate to describe real ballistic propagation~\cite{Vainstein05}  than Eq.~(\ref{N1}), see Sec.~\ref{sec.ballistic}.


\section{Reversibility and Correlation functions}
\label{sec.reversibility}

As early as 1876, Loschmidt called attention to the reversibility paradox~\cite{Loschmidt76}. His paradox states that all molecular processes must be reversible, since there is a symmetry between past and future $t \rightarrow -t$ in the laws of physics. 
Consequently, statistical mechanics must be reversible, in apparent contradiction with thermodynamics, where certain processes are irreversible.
At that time, reversible physics was composed of classical mechanics and electrodynamics. This paradox, together with the dynamical problems of Liouville, Zermello, and Poincar\'e are central in the work of Boltzmann. Those lead to the Boltzmann equation, to the H theorem~\cite{Huang87}, and to the studies of the Poincar\'e recurrence theorems. 

Again, our aim here is not to go into extensive mathematical proof; rather we focus on the correlation function.
In the definition of the correlation function we have used
\begin{equation}
\label{tt}
R(t_1-t_2)=\frac{\langle A(t_1)A(t_2)\rangle}{\langle A^2\rangle}
\end{equation}
and
\begin{equation}
\label{tr}
R(-t)= R(t).
\end{equation}
The first relation relies on our basic knowledge of the temporal invariance of physical laws.
 However, if the evaluation is made far from equilibrium, the correlation function $R$ may depend both on $t_1$ and $t_2$, and not on their difference.

The second relation is  time reversal, which can be easily understood for a classical variable where $A(t)$ and $A(0)$ commute. Given a string  of values $A(t_i)$ $i={1,2,\ldots,N_{int}}$, for large $N_{int}$,  one can obtain the relations given by Eqs.~(\ref{tt}) and (\ref{tr}). For quantum systems, the reader is recommended the review of Balucani \emph{et al.}~\cite{Balucani03}   

A great achievement in the discussion of time reversal symmetry for macroscopic systems was made by Onsager. 
He considered a solid subject to a general field $E$. In the linear regime, the field induces a generalized current density $J$ of the form
\begin{equation}
 J_i= \sigma_{i,k}E_k.
\end{equation}
The susceptibility $\sigma$ satisfies
\begin{equation}
\label{Onsager}
   \sigma_{i,k}=\mu_{i,k}\sigma_{k,i}.
\end{equation}
Here, $\mu_{i,i}=1$ for the diagonal terms. The off-diagonal terms are $\mu_{i,k}=1$ for variables which do not change sign under time reversal, such as the electric field, and $\mu_{i,j}=-1$ for variables which do change, such as the magnetic field. Equation~(\ref{Onsager}) is the Onsager reciprocal relation.

A natural generalization of susceptibility is the correlation function or the response function, sometimes called a Green function. Consider, for example, that one applies a perturbation $P(x_1)$ at the position $x_1$ and  wants to know the disturbance $S(x_2)$ at $x_2$. In the linear regime, we get
\begin{equation}
S_i(x_2)= \int G_{i,k}(x_2,x_1)P_k(x_1)dx_1.
\end{equation}
   For systems with translational invariance, we expect that 
\begin{equation}
\label{si}
G(x_1,x_2)= G(x_1-x_2).
\end{equation}
When the translational invariance  is broken, due to the existence of surfaces such as in a film~\cite{Oliveira81,Oliveira93}, defects, or topological disorder~\cite{Moraes03}, the response becomes a function of both variables $x_1$ and $x_2$. 
However, Oliveira~\cite{Oliveira81} has given a proof based on time reversal symmetry that
\begin{equation}
\label{ss}
 G_{i,k}(x_1,x_2)=  \mu_{k,i} G_{k,i}(x_2,x_1)
\end{equation}
The tensor $\mu_{i,j}$ here is the same as the one in  Eq.~(\ref{Onsager}), i.e., we lose the space invariance, but, on the other hand, we gain a useful space exchange symmetry.
 This result is general and has many applications. For example, it was used to explain certain asymmetries found in light scattering~\cite{Oliveira81}.                                                                                

The usual Onsager reciprocal relations are actually limited to systems asymptotically close to equilibrium. For example, they apply at the level of the Navier-Stokes equations for a simple fluid, but fail for the higher-order corrections to those equations, as pointed out by McLeannan~\cite{McLeannan74,Dufty87}. Dufty and Rub\'{\i}~\cite{Dufty87} generalized Mcleannan's work to nonequilibrium stationary  states.

Many correlation functions of the form $\langle A(x_1,t_1)A(x_2,t_2) \rangle$ have properties similar to Eq.~(\ref{si}). Notice that even in nonlinear systems, sometimes it is possible to make some general statements.
 For example, in the growth process, the height of a surface $h(x,t)$ is a function of the position $x$ and of the time $t$. The main studied quantity is the roughness~\cite{Barabasi95}, defined by the mean square fluctuation 
\begin{equation}
\Delta h^2(x,t)= \langle (h(x,t)-\langle h(x,t)\rangle)^2\rangle.
\end{equation}
The roughness satisfies the scaling laws
\begin{equation}
b\Delta h^2(bx,b^zt)= \Delta h^2(x,t).
\end{equation}
where $b$ is a number and $z$ is the growth exponent. Notice that this relation holds only statistically. Many  symmetries or scaling in the correlation function  hold in situations where nothing can be said for a single process.

We now return to the correlation functions of the GLE. Note that the memory is an even function of $t$, independent of the NDS (see Eq.~(\ref{Gamma})). 
The analytical extension of the Laplace transform of an even function is an odd function, $\widetilde{\Pi}(-z)=-\widetilde{\Pi}(z)$. Consequently from Eq.~(\ref{Rz}), $\widetilde{R}(-z)=-\widetilde{R}(z)$, and, by a converse argument, $R(t)$ is an even function~\cite{Vainstein05a}.
 This is in agreement with the results by Lee for Hamiltonian systems~\cite{Lee83}. Notice also that Eq.~(\ref{GR}) requires the derivative of $R(t)$ to be null at the origin. Since both memory and $R(t)$ are even, they can be written as 

\begin{equation}
\label{G2}
\Pi (t)=\sum _{n=0}^{\infty }b_{n}t^{2n},
\end{equation} 
and
\begin{equation}
\label{R2}
R(t)=\sum _{n=0}^{\infty }a_{n}t^{2n}.
\end{equation}

Unfortunately, a large number of works has been presented in the literature where the correlation function does not satisfy these requirements. We shall not comment further on these works here: some may be useful approximations, others represent artificial solutions. The reader should be cautious in identifying them.

Exponentials, stretched exponentials, and power laws are examples of asymptotic behavior that can be obtained from  more complex even functions~\cite{Vainstein05a}, but obviously they do not fulfil Eq.~(\ref{GR}).

Determination of the coefficients in Eqs.~(\ref{G2}) and (\ref{R2}) can be done for every specific noise. For short times, those equations yield  $R(t)= \cos(\omega_0t)$, where $\omega_0=\sqrt{\Pi(0)}$. For broadband noise, the asymptotic times yield exponential decay for normal diffusion. For anomalous diffusion, the behavior is of a stretched exponential followed by inverse power-law.  For short band noise a very rich oscillatory behavior may be found~\cite{Vainstein05a}.

A large and growing literature in which non-{ex\-po\-nen\-tial} behavior  has been observed for correlation functions can be found in the following articles: in glasses and supercooled liquids~\cite{Xia01}, frustrated lattice gases~\cite{Vainstein03a}, liquid crystals~\cite{Benmouna01,Santos00,Licinio02}, plasmas~\cite{Ferreira91}, hydrated proteins~\cite{Peyrard01}, growth~\cite{Colaiori01} and disordered vortex lattice in superconductors~\cite{Bouchaud91}.                                                  

\section{Mixing, Ergodicity, and the Fluctuation-Dissipation Theorem}
\label{sec.violation}

In this section, we arrive at the central point of this work by showing that  the EH, Eq.~(\ref{EH}), the MC, Eq.~(\ref{MC}), and the second FDT, Eq.~(\ref{FDT}) are strongly connected in the GLE. 
Consequently, one could expect that the violation of one of these  conditions could lead to the violation of the others. However, we will show that there is a hierarchy among the three concepts, in such a way that some may be violated, while others are not. Most of the systems that present violation of the FDT are complex, such as supercooled organic liquids, some algebraic maps, evolutionary models, and the eternal spin-glass problem. We try to show here a minimal condition for violation of that hierarchy.

We may expect from Eqs.~(\ref{GLE}) and (\ref{FDT}) that
a system will be driven to an equilibrium state, i.e.

\begin{equation}
\lim _{t\rightarrow \infty } \overline{A^{2}(t)}=\langle A^{2}\rangle_{eq},
\label{A0}
\end{equation}
 which can be identified with the EH. We shall see, however, that this is not always the case for superdiffusive dynamics.
 
Note that the Laplace transform of Eq.~(\ref{GLE}) suggests a solution of the form
\begin{equation}
A(t)= A(0)R(t)+\int _{0}^{t}R(t-t')F(t')dt',
\label{A1}
\end{equation}
 where we have an ensemble of initial $A(0)$. 
Squaring Eq.~(\ref{A1}) and taking the ensemble average, we obtain for the asymptotic behavior~\cite{Costa03} 

\begin{equation}
\langle A^{2}(t\rightarrow \infty) \rangle =\langle A^2\rangle_{eq}+R^2(t\rightarrow \infty)[\langle  A^2(0)\rangle -\langle  A^2\rangle_{eq}].
\label{A2}
\end{equation}
This simple result leads to very important consequences. First, the system will reach full equilibrium only if the MC, Eq.~(\ref{MC}), holds.  Second, the EH holds if the MC holds.
 Finally, the FDT will hold only if the EH holds. Consequently, the FDT is the end  validation of the sequence: Mixing $\Rightarrow$ Ergodic Hypothesis $\Rightarrow $ Fluctuation-Dissipation Theorem. 
Observe that if the MC is violated, then  the final value of  Eq.~(\ref{A2}) will depend on the initial conditions. That is just the essence of the MC.

At this point,  we shall call again the attention to Lee's work in ergodicity~\cite{Lee01}. Unlike any other previous attempt at establishing the validity of the Boltzmann EH, his work approaches the time average directly and explicitly, which was  made possible by his recurrence relation method~\cite{Lee83a}.
 Moreover, it has been demonstrated in several exact solvable  models when the hypothesis is valid. When it is not valid, it is shown the reason  why the hypothesis fails.

If the mean square value of $A$ can be associated with a given temperature by the equipartition theorem, we have $\langle  A^2(0) \rangle \sim T_0$ for the initial temperature, 
$T$ for the reservoir temperature, and $T_{eff}$ for the final effective temperature. Equation~(\ref{A2}) becomes
\begin{equation}
T_{eff} =T +  R^2(t\rightarrow \infty)[T_0-T].
\label{Teff}
\end{equation}
From Eq.~(\ref{Rlim}), we see that the MC is satisfied for $0<\alpha<2$. For the ballistic motion, $\lim_{z\rightarrow 0}\Pi(z)= az$, $\nu=1$, $\alpha=2$, and  $R(t\rightarrow \infty)= (1+a)^{-1}$.
This system never thermalizes to the reservoir temperature, unless it  already starts at equilibrium.
 The system acquires an effective temperature different from that of the reservoir. This effective temperature is a signature
 of metastability found in glasses, where the FDT does not hold~\cite{Kauzmann48,Santamaria-Holek01,Ricci-Tersenghi00,Costa03,Costa04}.

The first observation of such phenomena was reported by Kauzmann~\cite{Kauzmann48}.
He noticed that when the entropy of a supercooled liquid is extrapolated below the glass temperature $T_{g}$, it can become smaller than the entropy of the cryatallline solid. To avoid
this paradox, he suggested the existence of an  effective spinodal temperature $T_{sp}$ in the supercooled liquid phase.
Ricci-Tersenghi \emph{et al.}~\cite{Ricci-Tersenghi00}
and Cavagna \emph{et al.}~\cite{Cavagna03} performed single-spin-flip Monte Carlo simulations in square lattices with frustration, in which  they obtained effective temperatures $T_{eff}\neq T$. Methods for measuring those effective temperature~\cite{Exartier00,Bellon05}, and many attempts to get a form of FDT for inhomogeneous systems have been discussed in the literature~\cite{Belyi02,Belyi04,Belyi04a}.

It has been shown that a drastic elimination of the fast degrees of freedom in the dynamics of a system may lead to a violation of the fluctuation-dissipation theorem~\cite{Vilar01}. 
This is due to the fact that equilibration in the coarsened description does not necessarily imply full equilibration of the system; therefore a fluctuation-dissipation relation, 
whose validity is limited to equilibrium or local equilibrium states~\cite{Callen52,Greene52,deGroot84}, may not exist.
The theorem is valid when there is a great disparity between slow and fast scales in such a way that faster scales relax practically immediately.
 This feature has been found in very different situations as in the diffusion of a Brownian particle in a shear
 flow~\cite{Rubi88,Santamaria-Holek01,Maroja01}, in the anomalous diffusion problem~\cite{Morgado02,Costa03,Costa04},
 in systems undergoing activated dynamics~\cite{Perez-Madrid03,Naspreda05}, and in slow relaxation of supercooled colloidal systems~\cite{Rubi04}.
This common scenario may suggest that the violation of the fluctuation-dissipation theorem could originate from the lack of ergodicity inherent to a coarsened description, which is related to the tacit reduction of the dimensionality of the system phase space.

\section{Ballistic Motion }
\label{sec.ballistic}
As discussed earlier in  Sec.~\ref{sec.memory}, most of the anomalous diffusion is subdiffusive, what can also be observed in most conductors~\cite{Dyre00}. 
However, very recently in the history of conductivity investigations, superdiffusive and even ballistic motion have been produced in laboratories. 
Indeed, we can find reliable reports on ballistic conductivity in carbon nanotubes~\cite{Frank98,Poncharal02}, in semiconductors~\cite{Hu95}, and in semiconductor superlattices with intentional correlated disorder~\cite{Bellani99,Bellani00}.
  
For a simple description of ballistic diffusion, we use  Eqs.~(\ref{Gamma}) and (\ref{N1}) and obtain

\begin{equation}
\Pi (t)=\frac{2\gamma_0}{\pi} \left[\frac{\sin (\omega _{2}t)}{t}-\frac{\sin (\omega _{1}t)}{t}\right].
\label{14}
\end{equation}
 The Laplace transform of Eq.~(\ref{14}) gives, as $z\rightarrow 0$, 
$\widetilde{\Pi }(z)\sim z$. Consequently, $\nu =1$ and
$\alpha =2$, which is the ballistic limit.
 If we set $\gamma_0 =\pi\omega _{2}/4$, the initial temperature $T_0=0$, 
in Eq.~(\ref{Teff}),
we get the effective temperature as $T_{eff}$ as 
\begin{equation}
\lambda^*=\frac{T_{eff}}{T}=1-\left(\frac{2\omega _{1}}{\omega _{1}+\omega _{2}}\right)^{2}.
\label{15}
\end{equation}
 Equation~(\ref{15}) has a control  parameter  $\omega _{1}$, which
measures the ``hole'' in the density of states, and how far we
are from the result predicted by the FDT.

\begin{figure}
\centering
\includegraphics[height=7cm,width=5cm,angle=270]{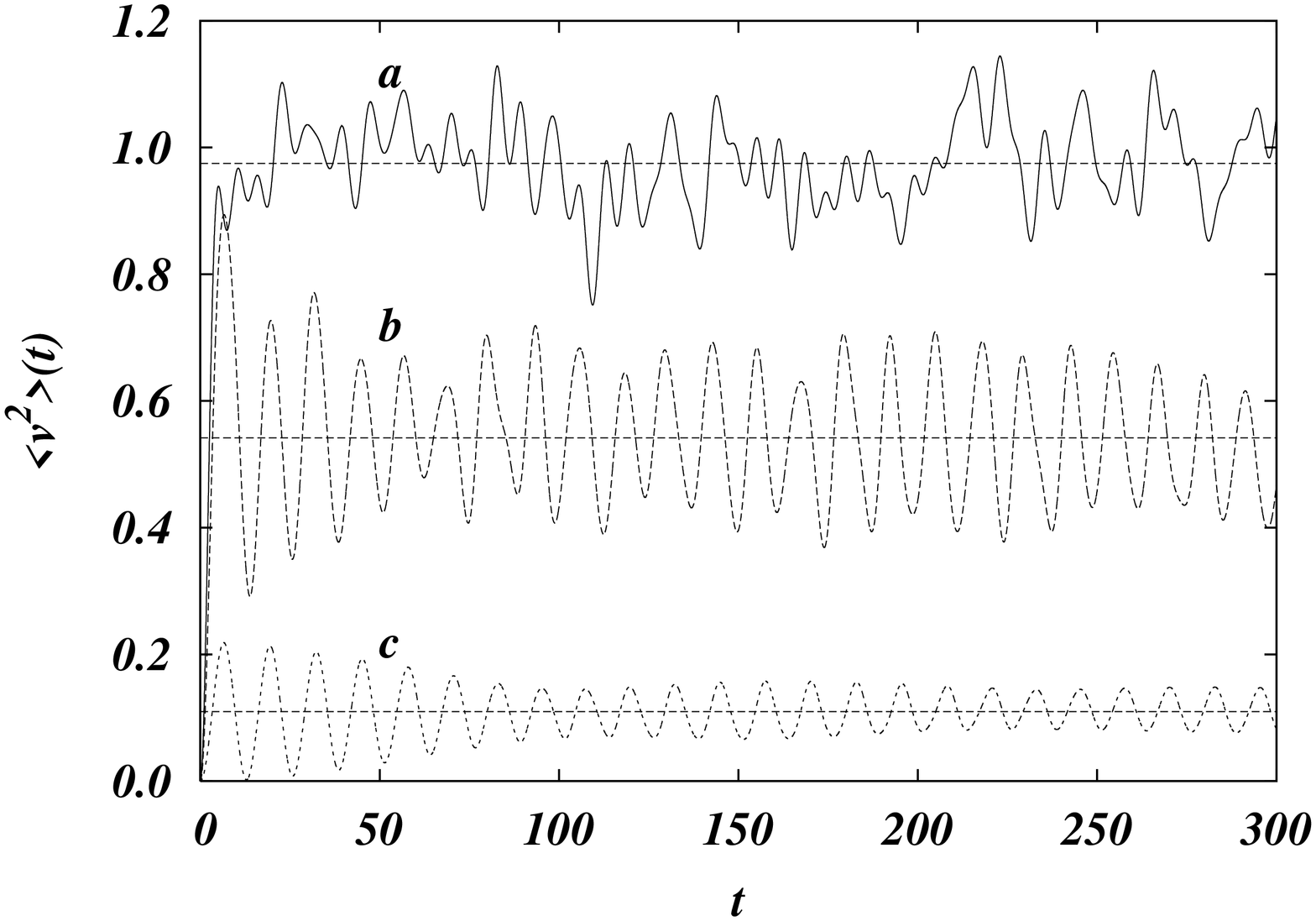}
\caption{Normalized mean square velocity as a function of time for the memory
given by Eq.~(\ref{14}). Here $\beta =\omega _{2}/2$ and $\omega _{2}=0.5$.
Each curve corresponds to a different value of $\omega _{1}$. a)
$\omega _{1}=0$; b) $\omega _{1}=0.25$; c) $\omega _{1}=0.45$.
The horizontal lines correspond to the final average value $\lambda _{s}$.
In agreement with the theoretical prediction, $\lambda_s$ decreases
as $\omega _{1}$ grows.}
\label{fig:1} 
\end{figure}

Now, we examine the case when $A(t)=v(t)$, the particle's velocity, so that we obtain $\langle v^{2}(t)>=\langle v^{2}\rangle_{eq}\lambda (t)$.
We simulate the GLE for a set of $10000$ particles starting at rest at the origin, using the memory in Eq.~(\ref{14}) with $\omega _{2}=0.5$
and different values of $\omega _{1}$. The results of these simulations
are shown in Fig.~\ref{fig:1}, where we plot $\langle v^{2}(t)\rangle$. We used the normalization
$\langle v^{2}\rangle_{eq}=1$, so that $\langle v^{2}(t)\rangle=\lambda (t)$. Notice that
$\lambda (t)$ does not reach a stationary value; rather, it oscillates
around a final average value $\lambda _{s}$. This value of $\lambda _{s}$
should be compared with $\lambda ^{*}$ obtained from Eq.~(\ref{15}).

\begin{figure}
\centering
\includegraphics[height=7cm,width=5cm,angle=270]{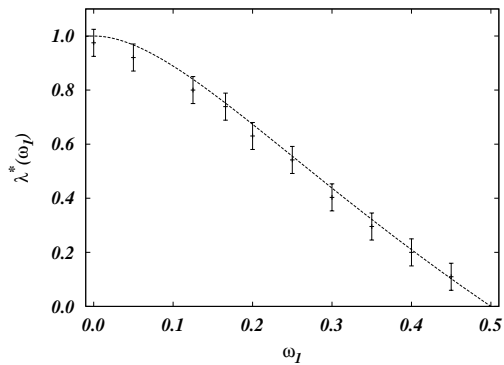}
\caption{$\lambda ^{*}$ as a function of the parameter $w_{1}$. The line corresponds to the theoretical
prediction given by Eq.~(\ref{15}). Each point corresponds to a value of $\lambda _{s}$ obtained from simulations like those described in Fig.~\ref{fig:1}.}
\label{fig:2} 
\end{figure}

In Fig.~\ref{fig:2}, we plot $\lambda ^{*}$ as a function of $\omega _{1}$
as in Eq.~(\ref{15}) with a fixed value of $\omega _{2}=0.5$. We also plot the  average values $\lambda _{s}$ obtained from simulations for
different values of $\omega _{1}$. Notice that as $\omega _{1}$ increases, $\lambda ^{*}$ decreases as expected. The agreement between simulations and Eq.~(\ref{15}) shows that we can predict the average value $\lambda _{s}$, even when the FDT does not hold.

\section{Shape of things to come}
\label{sec.shape}

After  the description of some ``well-established  physics'', we shall take this  section to discuss  some less conventional ideas about irreversibility, diffusion, fluctuations, and the approach to equilibrium.

 Blasone \emph{et al.}~\cite{Blasone01} have shown that a quantum harmonic oscillator can emerge from a couple of classical harmonic oscillators. Every classical oscillator will obey the laws of classical physics, however, together their behavior will follow quantum mechanics.
Bir\'o \emph{et al.}~\cite{Biro01}, studying the quantization of classical fields, demonstrated that a classical system that operates in 5 dimensions can transmute into a quantum system in 4 dimensions.

In the same context, the method of Lie symmetry applied to differential equations has been often invoked as a mechanism to derive, not only solutions, but also classes of Fokker-Planck equations with non-trivial drift and diffusion terms~\cite{Suzuki83,An84,Cicogna89,Shtelen89,Rudra90,Cicogna90,Spichak99,Cherkasenko95,Cardeal02}.
The study of Lie-group representations in space-time has been recently developed, following similar procedures as those of field theories~\cite{deMontigny03}. As a result, the usual Fokker-Planck equation has been derived from a $U(1)$ gauge invariant Lagrangian, and the generalization of such a formalism for the $SU(2)$ symmetry has provided  a new class of Fokker-Planck dynamics, which is non-abelian gauge invariant. The drift and the diffusion terms, in this situation, are associated with a tensor metric in a Riemannian manifold. This manifold is based on a Galilean metric space-time,  introduced (via the light-cone of a (4+1) Minkowski-like space) to derive the non-relativistic physics in a covariant fashion~\cite{Duval85,Takahashi88,deMontigny01}.

The apparent contradiction between the irreversibility of the macroscopic phenomena and  the time-reversal symmetry of the fundamental laws has driven passionate discussion since Boltzmann times. This paradox of irreversibility, discussed in Sec.~\ref{sec.reversibility},  finds its place both in classrooms and in highly specialized conferences.  According to Zwanzig~\cite{Zwanzig01}, there is no paradox. According to Chaves \emph{et al.}~\cite{Chaves94}, the paradox still remains as a problem far from being solved.
Though some people believe that the only mystery related to irreversibility is the fact that the universe started in a very special initial state, the question is in fact much subtler and deep. The point is that the laws of quantum mechanics warrant that time evolution  of an isolated system is described by a unitary operator that keeps constant the value of the entropy. The only escape from that fate could be quantum gravity, a theory still to be constructed. The point is that the gravitational interaction has infinite quantities that have not been renormalized. Thus, we cannot assure that the zero-point quantum fluctuations of the gravitational field -- which in fact are fluctuations of  space-time itself! -- result only  in the renormalization of the physically observed quantities.                    

Those fluctuations can in fact create a non-unitary contribution to the quantum mechanics evolution operator and thus be a fundamental source of irreversibility.
Chaves \emph{et al.}~\cite{Chaves94} suggested that this in fact occurs. On the basis of the quantum fluctuations of the metric tensor, they proposed an extra term in the Schr\"odinger equation which makes time evolution operator non-unitary.
Their calculations demonstrated that the coherence time of the microscopic system would be too long to be observed, but macroscopic systems would decohere very quickly.
Acebal \emph{et al.}~\cite{Acebal98} demonstrated that those metric fluctuations could also remove the infinities that plague quantum field theories.

 These unconventional analyses may prove useful for other  derivations. They require a  deeper understanding of time and space in the field of statistical mechanics.

\section{Spatio-temporal conjecture for disordered system}
\label{sec.conjecture}

Up to now, we have discussed stochastic systems, i.e., systems with noise that we shall name \emph{temporal disorder}.
For those, given the NDS, $\rho(\omega)$, it is possible to obtain the memory and then, by using Eq.~(\ref{alpha}), the diffusive exponent $\alpha$.

A second class of systems is composed by those which present \emph{spatial disorder}. They have been thoroughly investigated in the last half century~\cite{Anderson58}, nevertheless, some questions concerning localization or diffusion still remain open. Let us consider, as an example, the Heisenberg chain~\cite{Evangelou92,Moura02}

 \begin{equation}
\label{Heisenberg}
H=-\sum _{l=1}^{N}J_{l}\mathbf{S}_{l}\cdot \mathbf{S}_{l+1},
\end{equation}
 where $S=1/2$. Here, $J_l$ is the exchange integral at the site $l$. Equivalently, we could consider the disordered harmonic chain~\cite{Moura03}  or even the Anderson model~\cite{Moura98}.

   Can we predict the properties of those systems in the same way we do for the GLE? The answer is partially yes, partially no. The conjecture~\cite{Vainstein05}, being valid, will help to answer those questions.

  Consider a system which presents fluctuations in its energy density of states $D(E)$; let us call $\rho_F(E)$ the fluctuation density, then
\begin{equation}
\rho_F(E) = \rho(E).
\label{conjecture}
\end{equation}
If this is true, then  $\rho_F$ can be introduced in Eq.~(\ref{Gamma}) to obtain the diffusive exponent $\alpha$.
This is the spatio-temporal conjecture~\cite{Vainstein05}. This conjecture has been verified for the quantum disordered Heisenberg chain~\cite{Vainstein03} and is under consideration for many similar systems.

\section{Conclusion}
\label{sec.conclusion}

In this review we discussed some old and permanent problems of statistical mechanics. 
The way a system approaches equilibrium is connected with some basic questions in physics such as the reversibility paradox, the mixing condition, the ergodic hypothesis and the fluctuation-dissipation theorem. We have drawn a line between them and, in particular, we have discussed the hierarchy  MC $\Rightarrow$ EH $\Rightarrow$ FDT, established by Costa \emph{et al.}~\cite{Costa03}. 

We have approached the problem taking diffusion as a main phenomena in physics, since most processes are related with transport of matter, energy, or information.
In this context, the validity of the FDT is exhibited for ballistic motion.
Ballistic motion is presented here as the frontier between a stochastic
process described by a GLE and other processes such as hydrodynamical ones.

We discussed relaxation processes and the conditions that the correlation functions must fulfill.
We presented, in Sec.~\ref{sec.shape}, discussions on the frontier of physics with particular consequence to statistical physics. We revive the reversibility paradox as an unclosed subject, as well as the  MC, EH and FDT.  We discussed a conjecture that, if valid, will make an important connection between stochastic and Hamiltonian descriptions. 

Nonlinear dynamics is a field which deserves much attention; in particular, the coalescence of trajectories has been intensively studied in the last few years~\cite{Longa96,Boccaletti02}. There, the restriction of the degrees of freedom may confirm as well the hierarchy exposed here.

We have not focused deeper on real complex systems; we chose to follow easy-to-understand concepts where limits could be analytically obtained. This gave us a good framework for analyzing more complex structures.

  We also tried to show that  a given result may be obtained through many different formalisms. 
 Feynman once said: ``A physicist must know at least five different ways to obtain a result''~\cite{Feynman94}. If we consider the FPE, the FFPE, and the GLE as alternative approaches, we are close to fulfilling Feynman's requirement. It is nice to know that those approaches agree in the main results; however, the full picture has not yet been drawn, particularly for anomalous diffusion.
 
 Although anomalous diffusion remains as a surprising phenomenon,
we hope that this work will help in the centennial effort to understand diffusion and the relation between fluctuation and dissipation. A generalization of the FDT to include nonlinearities and ballistic motion is necessary, what will require a deeper understanding of systems far from equilibrium.
\vspace{0.5cm}

ACKNOWLEDGEMENTS:
One of us (FAO) would like to thank the hospitality and enlightening
discussions with professors Alaor Chaves (Belo Horizonte), Alex Hansen (Trondheim), Fernando Moraes (Jo\~ao Pessoa), Howard Lee (Athenas-USA), Jo\~ao Florencio (Niter\'oi), Jon Otto Fossum (Trondheim), Lech Longa (Krakow), Miguel Rub\'{\i} (Barcelona), Nitant Kenkre (New Mexico), Robin Stinthcombe (Oxford), Sam Edwards (Cambridge), and Silvio Salinas (S\~ao Paulo).
In Bras\'{\i}lia, we would like to thank  Ademir Santana, Adriano Batista, Annibal Figueiredo, Geraldo Magela, Geraldo Silva, Hugo Nazareno, Rafael
Morgado, and  Sebasti\~ao Silva.  This work was supported by CAPES, CNPq, FINEP, and FINATEC.


\bibliographystyle{unsrt}
%


\printindex

\end{document}